# Non-Linear Cluster Lens Reconstruction

Nick Kaiser

*Canadian Institute for Advanced Research and*
*Canadian Institute for Theoretical Astrophysics, University of Toronto*
*60 St. George St., Toronto, Ontario, M5S 1A7*

## ABSTRACT

We develop a method for general non-linear cluster lens reconstruction using the observable distortion of background galaxies. The distortion measures the combination $\gamma/(1-\kappa)$ of shear $\gamma$ and surface density $\kappa$. From this we obtain an expression for the gradient of $\log(1-\kappa)$ in terms of directly measurable quantities. This allows one to reconstruct $1-\kappa$ up to an arbitrary constant multiplier. Recent work has emphasised an ambiguity in the relation between the distortion and $\gamma/(1-\kappa)$. Here we show that the functional relation depends only on the parity of the images, so if one has data extending to large radii, and if the critical lines can be visually identified (as lines along which the distortion diverges), this ambiguity is resolved. Moreover, we show that for a generic 2-dimensional lens it is possible to locally determine the parity from the distortion. The arbitrary multiplier, which may in fact take a different value in each region bounded by the contour $\kappa = 1$, can be determined by requiring that the mean surface excess vanish at large radii and that the gradient of $\kappa$ should be continuous across $\kappa = 1$. We show how these ideas might be implemented to reconstruct the surface density, if necessary without use of the data in regions where determination of the parity is insecure, and we show how one can measure the mass contained within an aperture, again, if necessary, without using data within the aperture.

*Subject headings:* cosmology: observations – dark matter – gravitational lensing – galaxy clusters

Observations of giant arcs, arclets and weak distortion at large radii show galaxy clusters to be potent gravitational lenses. With deeper observations of very massive clusters it is quite foreseeable that we will soon have quite detailed maps of the distortion field in the best cases, and in both the linear and non-linear regime. Here we will address the question: Given only the distortion of background objects, how can one reconstruct the surface mass density in the lens? The observed surface brightness for an object seen through a lens is $f_{\rm obs}(\theta_i) = f_{\rm true}(\psi_{ij}\theta_i)$. The angular position is measured relative to some arbitrary fiducial point on the object and where the magnification matrix is $\psi_{ij} = 1 - \phi_{,ij}$ where the surface potential $\phi$ is related to the surface density by $\nabla^2 \phi = 2\Sigma/\Sigma_{\rm crit} \equiv 2\kappa$. For a flat, critical density universe $\Sigma_{\rm crit} = (4\pi a_l w_l(1 - w_l/w_s))^{-1}$, where $w \equiv 1 - 1/\sqrt{1+z}$. We will assume for simplicity that the sources are much more distant than the lens ($w_s \gg w_l$) so the distortion is effectively independent of source distance. This is not really a limitation on the method; when this inequality is not strongly satisfied we actually have *more* information at our disposal, but this can be reduced, in principle, to give the distortion on a single plane.

There are many possible statistics which can be used to measure the distortion; typically these involve measuring quadrupole moments and forming some 2-component measure of the polarisation of the image shapes. The application of a locally constant shear will distort the distribution of observed polarisations from its intrinsically isotropic form, so with sufficiently numerous background galaxies one can measure the distortion strength with high precision. The best one can hope to learn from such studies is the orientation and the ratio of the absolute values of the eigenvalues of $\psi_{ij}$. This is illustrated in figure 1 which shows the distortion of intrinsically circular objects by a simple model lens. For concreteness let $\varphi_0$ be the angle of the major axis relative to the $x$-axis and let $R \leq 1$ be the ratio of the short to the long axis.

Let us now assume that we are supplied with sufficiently precise and detailed maps of $R(\vec{\theta})$ and $\varphi_0(\vec{\theta})$. Let the eigenvalues of $\phi_{,ij}$ be ordered such that $\lambda_1 > \lambda_2$ and let the $x_1$ direction lie at an angle $\varphi_\phi$ relative to the cartesian $x$-axis. In the diagonal frame we have $\psi_{11} = 1 - \kappa - \gamma$ and $\psi_{22} = 1 - \kappa + \gamma$

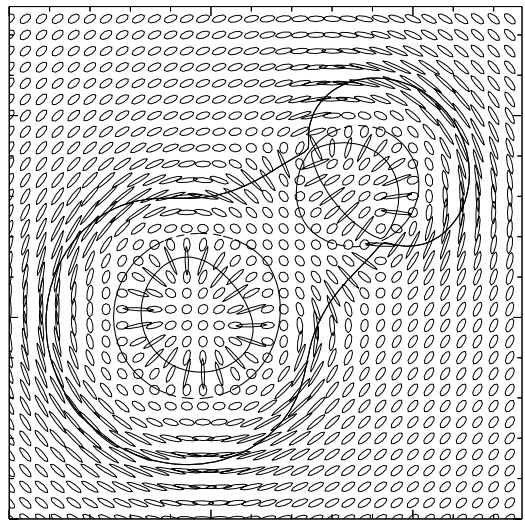

Fig. 1.— Distortion field for simple "dumbbell" cluster lens model. At each grid point on the deflector plane we have drawn the apparent shape for an intrinsically circular background object. The critical lines are shown (solid). The distortion diverges on these lines (though note that the arcs are not, in general, aligned parallel to the critical lines). The critical density line is shown dashed. The distortion vanishes along this line.

where $\gamma \equiv (\lambda_1 - \lambda_2)/2$ is the shear, so

$$\frac{\gamma}{1-\kappa} = \frac{1 - \psi_{11}/\psi_{22}}{1 + \psi_{11}/\psi_{22}} \quad (1)$$

The fact that the observable eigenvalue ratios only determine this combination of $\gamma$ and $\kappa$ was emphasised by Schneider and Seitz (1994). They showed that this results in a 'global invariance transformation'; there is a one parameter family of surface density configurations which are compatible with any given distortion pattern, and we will recover this below.

Schneider and Seitz also noted an ambiguity in the relation between $\gamma/(1-\kappa)$ and the observables $R$, $\varphi_0$; clearly this depends on the signs of the eigenvalues of $\psi_{ij}$. Consider first the case where $\psi_{11}$, $\psi_{22}$ have the same sign (even parity image). If they are both positive the smallest eigenvalue is $\psi_{11}$, so $R = \psi_{11}/\psi_{22}$, and the long axis is then aligned with the $x_1$ direction, so $\varphi_\phi = \varphi_0$. If on the other hand they are both negative, the smallest absolute eigenvalue is now $\psi_{22}$, so $R = \psi_{22}/\psi_{11}$ and the long axis is perpendicular to the $x_1$ direction so $\varphi_\phi = \varphi_0 + \pi/2$. These appear to give different functions for $\gamma/(1-\kappa)$ in terms of the observables $R$, $\varphi_0$. However, this is in the diagonal frame. In the general coordinate frame we have $\gamma_1 = \gamma \cos 2\varphi_\phi$, $\gamma_2 = \gamma \sin 2\varphi_\phi$ and the numerical value of these components for given $R$, $\varphi_0$ are the same in both cases, so we have

$$\frac{\gamma_i}{1-\kappa} = e_i \qquad \text{even parity} \quad (2)$$

with $e_1 \equiv e \cos 2\varphi_0$, $e_2 \equiv e \sin 2\varphi_0$ and with $e \equiv (1-R)/(1+R)$. The boundaries of the even parity regions are the 'outer' critical line where $\psi_{11}$ vanishes and which lies at $\kappa \leq 1$, and the 'inner' critical line where $\psi_{22} = 0$ and which lies at $\kappa \geq 1$.

Now consider the case that the eigenvalues have opposite sign (odd parity image) i.e. $\psi_{11} < 0$. As before, the orientation depends on which eigenvalue is larger. They are equal and opposite along the line $\kappa = 1$ where the distortion vanishes. (Note that a closed loop on which the distortion vanishes can only occur in the odd parity region; in the even parity regime the distortion will generally only vanish at the singular 'umbilic' points where both eigenvalues are equal (Berry and Hannay, 1977).) If $\kappa < 1$ then $R = -\psi_{11}/\psi_{22}$ and $\varphi_\phi = \varphi_0$. Otherwise $R = -\psi_{22}/\psi_{11}$ and $\varphi_\phi = \varphi_0 + \pi/2$ and in either case we find

$$\frac{\gamma_i}{1-\kappa} = \frac{e_i}{e^2} \qquad \text{odd parity} \quad (3)$$

Thus in either case, the quantity we can hope to measure is $\gamma/(1-\kappa)$; in the even parity case it is equal to the 'ellipticity' parameter $e \leq 1$, and in the odd case it is equal to the inverse ellipticity. Given perfect observations one could simply identify the critical lines as lines connecting strongly distorted galaxies and, provided the data extend to large radii where it is safe to assume the parity is even, one could then unambiguously determine $\gamma/(1-\kappa)$. As we will shortly see, for a general two-dimensional lens one can also directly determine the parity from local observations of the distortion.

We would now like to eliminate the shear from equations 2, 3 to obtain an expression for $\kappa$ in terms of observables. Unfortunately there is no unique relation between $\kappa$ and $\gamma$, but there is a simple relation between the gradients of $\kappa$ and of $\gamma$: Since $\gamma_i = \{(\phi_{,11} - \phi_{,22})/2, \phi_{,12}\}$ we find $\partial \gamma_1/\partial x_1 + \partial \gamma_2/\partial x_2 = \phi_{,111}/2 - \phi_{,221}/2 + \phi_{,122} = \partial \kappa/\partial x_1$. Similarly, $\partial \gamma_2/\partial x_1 - \partial \gamma_1/\partial x_2 = \partial \kappa/\partial x_2$ or

$$\partial_i \kappa = D_{ij} \gamma_j \quad (4)$$

where $\partial_i \equiv \partial/\partial x_i$ and where we have defined the operator

$$D_{ij} = \begin{bmatrix} \partial_1 & \partial_2 \\ -\partial_2 & \partial_1 \end{bmatrix}. \quad (5)$$

It may seem somewhat mysterious that one can construct the vector $\partial_i \kappa$ out of $\gamma_i$ (which, while a 2-component entity, does not transform as a vector). The relation (4) relies on the fact that the shear field is derived from a single scalar function $\phi$, and thus is rather special. Replacing $\gamma_i$ in equation 4 by $(1-\kappa)e_i$ we have

$$\partial_i \kappa = (1-\kappa) D_{ij} e_j - e_j D_{ij} \kappa \quad (6)$$

or equivalently

$$\partial_i \log(1-\kappa) = -M_{ij}^{-1} D_{ij} e_m \equiv u_i \quad (7)$$

where

$$M_{ij} = \begin{bmatrix} 1 + e_1 & e_2 \\ e_2 & 1 - e_1 \end{bmatrix} \quad (8)$$

Similar expressions can be obtained for the odd parity case (or one can use these expression with $e \to 1/e$, though this will involve differentiating the inverse ellipticity which is probably not such a good idea).

Equation 7 is the main result of this letter; it gives the gradient of the scalar function $\log(1-\kappa)$ in terms of directly measurable quantities. With sufficiently detailed observations one can simply integrate this vector and reconstruct $1-\kappa$, though only up to an unknown multiplicative constant (this is Schneider and Seitz's global invariance transformation). In fact, the multiplicative factor can be set separately in each region bounded by the contour $\kappa = 1$, but this apparent ambiguity is removed by requiring continuity of $\vec{\nabla}\kappa$ across $\kappa = 1$.

We can now see how one can locally determine the parity: the vector $\vec{u}$ is the gradient of a scalar, so curl$u \equiv u_{1,2} - u_{2,1}$ should vanish, but will typically only do so if one is using the appropriate expression for $\gamma/(1-\kappa)$ in terms of $e$. This is again because the distortion $e_i(\vec{\theta})$, while apparently a two component object, has only a single scalar degree of freedom. This introduces special interrelations between the distortion values, and this will, in general, not be invariant under the transformation $e \to 1/e$. This is illustrated in figures 2,3 for our simple model lens.

With the parity determined our problem is formally solved: one can then determine the difference in $\log(1-\kappa)$ between any two points by performing a line integral. One could, for instance, estimate $\log(1-\kappa)$ at a point (relative to the very small average value around some boundary at large distance) by averaging over radial lines. In the limit that the shear is weak this becomes identical to the method of Kaiser and Squires (1993; hereafter KS), and provides a natural generalisation of the KS method to the non-linear situation.

It may be that secure identification of the critical lines may not be possible. In that case one might still be able to use the even parity curl-$\vec{u}$ statistic as a diagnostic to identify at least roughly the outer critical line, and one can simply modify the generalised KS statistic to avoid the use of data within the critical line (e.g. by averaging only over radial lines which do not cross the outer critical line). In this way one can establish the surface density $\kappa$ at all points outside the outer critical line. This is interesting, but one might be more interested in establishing the mass *within* the critical region or within some region with boundary outside the critical line. This can be done, and is explicitly independent of any data within the boundary, as we now show.

To determine the mass within some aperture we need to first determine the shear outside that aperture. To see this, consider an arbitrary circular loop and calculate the mean tangential shear (defined to be $\gamma_T = \gamma_1 \cos 2\varphi + \gamma_2 \sin 2\varphi$) around the loop: $\langle \gamma_T \rangle = \int d\varphi/(2\pi) \gamma_T$. If we construct local cartesian coordinates with radially directed component $r$ and azimuthal component $l$, we have $\gamma_T = (\phi_{,rr} - \phi_{,ll})/2 = \phi_{,rr} - \kappa$, so $\langle \gamma_T \rangle = (d/dr) \int d\varphi/(2\pi) \phi_{,r} - \langle \kappa \rangle$. The integral here is, by Gauss' law simply the mass enclosed within the loop divided by the loop circumference and we find

$$\langle \gamma_T \rangle = -\frac{1}{2} \frac{d\overline{\kappa}}{d \ln r} \qquad (9)$$

where $\overline{\kappa}$ is the mean surface density within the loop. If we perform the integral $\int_{r_1}^{r_2} d\ln r \langle \gamma_T \rangle$ (which may be expressed as a two-dimensional area integral and hence as a sum over discrete data values) we obtain the difference between the mean surface density in the disk $r < r_1$ relative to that in the control annulus $r_1 < r < r_2$ which provides a lower bound on the mass interior to $r_1$, tending rapidly to the true mass for large $r_2$, while using only data exterior to $r_1$. This statistic was used by Fahlman et al. 1994, and the idea is readily generalised to an aperture of arbitrary shape.

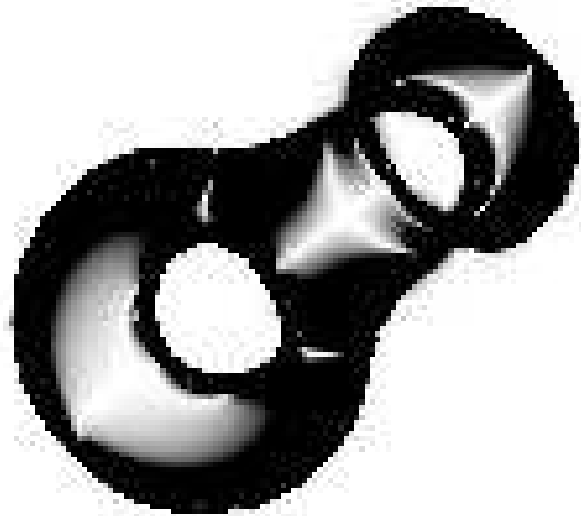

Fig. 2.— The ink density is proportional to the absolute value of curl-$\vec{u}$ calculated with even parity expression for $\vec{u}(R, \varphi_0)$. The curl vanishes outside the outer critical line and in the two regions which lie inside the inner critical line, and is non-zero elsewhere, and is. Curl-$\vec{u}$ is particularly large on the critical lines. The curl is implemented as a discrete differencing on a grid of $\vec{u}$ values, which themselves involve discrete differencing if the gridded $e_i$ values. This causes some smearing in the vicinity of the critical lines.

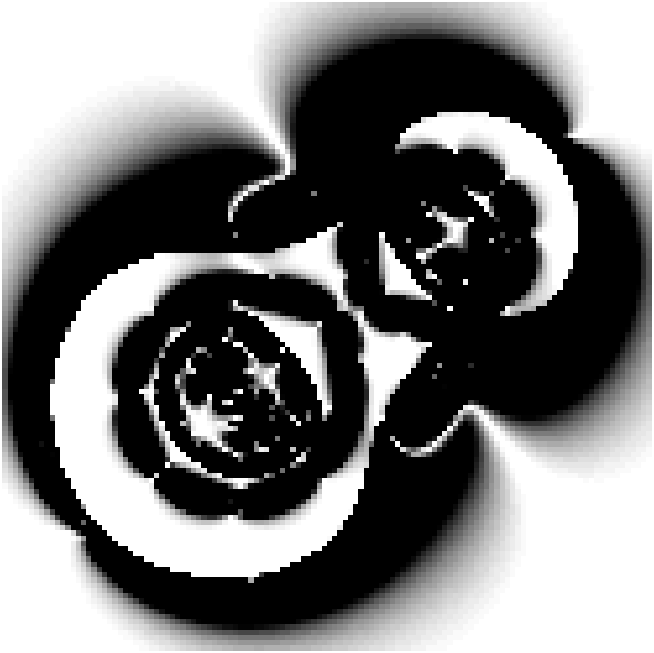

Fig. 3.— Curl $\vec{u}$ calculated with odd parity form for $\vec{u}(R, \varphi_0)$. The curl is non-zero in the even-parity regions. It is zero in most of the region lying between the critical lines (as it should be), but is also non-zero along the critical density contour where $\vec{u}$ diverges. Interestingly, a special case where the this method fails is a circular lens, where $\vec{u}$ is radially directed, though the curl-$u$ statistic still responds strongly to the presence of the critical lines.

How then do we determine the shear in our annulus? For a sufficiently compact lens $\kappa$ will be negligibly small in the annulus and we simply use $\gamma_i = e_i$ (and $\gamma_i = e_i/e^2$ in the odd parity region). In the general case we must first use $\vec{u}$ to calculate $\kappa(\vec{\theta})$ as outlined above (if the annulus were indeed empty we would find $\vec{u} = 0$ there), and then use $\gamma_i = (1 - \kappa)e_i$.

We have described a method which, given sufficiently precise data, allows one to reconstruct the full 2-dimensional surface density for an arbitrary lens from measurements of the distortion. The focus has been on the mathematics, rather than the 'engineering' problem of actually measuring the distortion field. Our main result is an explicit expression for the vector $\vec{u} \equiv \vec{\nabla} \log(1 - \kappa)$ in terms of observables, and we have shown how this provides a natural non-linear generalisation of the KS algorithm. In order to recover a detailed picture of the surface density in the very centre of super-critical lenses it will be necessary to locate the critical lines and apply the transformation $e \rightarrow 1/e$ to the data in the odd parity regime. The odd and even parity curl$\vec{u}$ statistics may prove helpful in this regard as they seem, for our idealised data at least, to cleanly detect these lines. It remains to be seen whether this will be practical with real data. A somewhat less ambitious approach is to use the even parity curl-$\vec{u}$ statistic to identify, at least approximately, the odd parity region, and we have shown how the generalised KS method may be implemented without using these data and we have also shown how the mass within some aperture lying outside the critical lines may be measured using only the exterior data.

---